\documentstyle[10pt,aaspp4]{article}
\begin{document}

\title{Gas-Rich Companions of Isolated Galaxies{\footnote{to appear in May 1999
Astronomical Journal}}}

\author{D.J. Pisano}
\affil{Dept. of Astronomy, University of Wisconsin}
\affil{475 N. Charter St., Madison, WI 53706}
\affil{Electronic Mail:  pisano@astro.wisc.edu}
\author{Eric M. Wilcots}
\affil{Dept. of Astronomy, University of Wisconsin}
\affil{475 N. Charter St., Madison, WI 53706}
\affil{Electronic Mail:  ewilcots@astro.wisc.edu}

\baselineskip1.1em
\begin{abstract}
We have used the VLA to search for gaseous remnants of the galaxy formation
process around six extremely isolated galaxies.  We found two distinct H~I 
clouds around each of two galaxies in our sample (UGC 9762 \& UGC 11124).  
These clouds are rotating and appear to have optical counterparts,
strongly implying that they are typical dwarf galaxies.  The companions are
currently weakly interacting with the primary galaxy, but look to be in 
stable orbits as they have dynamical friction timescales of 5-100 Gyr.  In
one case, (UGC 11124N), we see ongoing accretion of the companion on to the
primary galaxy.  The small mass ratio involved in this interaction means that
the resulting merger will be a minor one.  The companion does, however, contain
enough gas that the merger will represent a significant infusion of fuel to 
drive future star formation in UGC 11124 while building up the mass of the 
disk.

\end{abstract}

\section{Introduction}

How and when did galaxies form?  Current models of cold dark matter galaxy
formation suggest that spiral galaxies were built up via the accretion of
smaller bodies in a hierarchical merging process (e.g. Navarro, Frenk, \& White
1995).  Recent detections of H~I clouds near larger spiral 
galaxies may be signatures of this process occurring in the local universe.
Such detections include NGC 925, a late-type spiral (Pisano {\it{et al.}} 
1998); NGC 3432, NGC 4288, and NGC 4861 (Wilcots {\it{et al.}} 
1996); four of the sixteen low surface brightness (LSB) dwarfs, and four of 
the nine H~II galaxies 
observed by Taylor {\it{et al.}} (1993, 1996) also have small H~I companions.  
In all of these cases the companions contain 10$^7$-10$^8$M$_{\odot}$ of H~I 
amounting to 1-50\% of the mass of the primary galaxy.

Efforts to understand the origin of H~I clouds around galaxies are hindered by
the fact that the galaxies mentioned above were observed with no regard 
for their environment; the samples contain
quiescent and active galaxies in both sparse and dense environments.  As a 
result, the detection of these clouds provides only modest
insight into their origin.  Some of the aforementioned detections have been
interpreted as infalling gas from a primordial reservoir, as proposed for NGC 
628 by Kamphius \& Briggs (1991) and for IC 10 by Wilcots \& Miller (1998).  
On the other hand, galactic fountains and superwinds could be 
responsible for some of the high velocity clouds seen around large spirals 
(Wakker \& van Woerden 1997).  
Also, a number of the H~I clouds were found to be coincident
with faint companion galaxies (Taylor {\it{et al.}} 1993, Wilcots {\it{et al.}}
1996).  Still other detections could represent tidally stripped material from
an interaction (e.g. M81 group, Yun 1992).  Therefore, few, if any, of the H~I 
clouds represent unambiguous detections of the remnant reservoir of gas
from which the galaxies formed.

We have initiated a systematic search for the remnants of this reservoir of 
gas by concentrating on a sample of extremely isolated and quiescent galaxies.
By focusing on isolated galaxies, we greatly reduce the possibility that any
detections of H~I clouds are debris from recent interactions.  In addition, 
there is very little chance in such quiescent galaxies that superwinds or 
galactic fountains could expel such gas.  Here we report on the first results
from this search.  Section 2 discusses our sample selection and observations.
Section 3 summarizes our findings and the properties of the detected 
companions.  In section 4 we discuss the implications of these companions for
understanding galaxy formation and evolution; we summarize in section 5.

\section{Observations}

We have observed galaxies in Tully's ``Nearby Galaxies Catalog'' (1988) that 
were classified as completely isolated.  These galaxies are not in 
clusters, groups, or even associations.  According to Tully (1988), these 
galaxies have no known companions within 6 h$^{-1}$ Mpc.
The average galaxy density around these galaxies is less than 0.1 Mpc$^{-3}$ 
for galaxies with M$_B$ $\leq$-16 mag.
As a result, any H~I clouds detected are not likely to be the debris from a
tidal interaction.  We also selected out any galaxies classified as peculiar
so as to eliminate merger remnants from our sample and to insure we are not
biased towards detecting companions.  Furthermore, we are looking at quiescent
galaxies, hence minimizing the chances that the H~I clouds detected could have
been ejected from the primary galaxy.  Quiescence was determined by the
lack of references to starbursts or nuclear activity in the sample galaxies.  
This was done due to the lack of observations that could provide a better 
criterion for quiescence (such as a low H$\alpha$ luminosity).
In addition we chose galaxies with 
recession velocities between 1500 and 2600 km s$^{-1}$.  This places them 
between 23 and 35 Mpc for H$_o$=65 km/s/Mpc (to be used throughout this paper).
The near limit was selected such that a perturber traveling at 100 km
s$^{-1}$ could have interacted with the galaxy in the past Gigayear and still
be within the 30$^{\prime}$ beam of the VLA.  The far limit was chosen so that
we had at least two D-array beams ($\sim$60$^{\prime\prime}$) across a typical
galactic radius (20 kpc).  From this limited sample, we chose six galaxies at
random to observe with the VLA.

We observed the six galaxies with the Very Large Array{\footnote{The VLA is 
part of the National Radio Astronomy Observatory which is operated by 
Associated Universities Inc., under a cooperative agreement with the National 
Science Foundation}} in its D array configuration on 
November 15, 1997, and January 23, 1998.  Relevant observing parameters are
summarized in table 1.  We observed all galaxies with
5.2 km s$^{-1}$ velocity resolution in split IF mode providing almost 600 
km s$^{-1}$ of velocity coverage.  We spent approximately 2 hours on each 
galaxy, with observations of VLA primary (3C286 \& 3C48) and secondary 
calibrators interspersed with on-source observations resulting in a 1$\sigma$ sensitivity of 0.70 to 0.85 mJy/beam in each channel
corresponding to $\sim$10$^{18}$cm$^{-2}$ for each galaxy.  We flagged interference 
interactively, and the rest of the reduction proceeded in the usual manner 
using AIPS.  We made data cubes by combining the two IFs and 
using a natural weighting to restore the CLEAN'd circular beams with IMAGR
for all but one galaxy, NGC 3246. The
resulting beams have sizes between 61$^{\prime\prime}$ and 70$^{\prime\prime}$.  
For NGC 3246 we used a robust weighting scheme which provides a smaller
beam size of uniform weighting and nearly the sensitivity of natural 
weighting.  This yielded a beam size of 52$^{\prime\prime}$ and a 
1$\sigma$ sensitivity of 0.91 mJy/beam in each channel 
(1.9$\times$10$^{18}$cm$^{-2}$).  
While we made moment maps with blanking at the 2$\sigma$ level, all our
detections are above the 5$\sigma$ level.  

Dr. Liese van Zee kindly obtained B band images at the KPNO\footnote{Operated 
by the Association of Universities for Research in Astronomy, Inc. (AURA) under
a cooperative agreement with the National Science Foundation.} 0.9m telescope 
for the UGC 9762 and UGC 11124 fields.  She reduced the images in IRAF in the 
usual manner and registered to the J2000 coordinate grid.  All other optical
images were obtained from the Digital Sky Survey.

\section{Results}

Of the six galaxies observed, two (UGC 11124 \& UGC 9762, figures 1 \& 2) have 
distinct H~I-rich companions and one (NGC 3246 figure 3) has a somewhat 
lopsided disk.   The remaining three
galaxies have no companions down to our mass detection limit of 
$\sim$10$^7$M$_{\odot}$.  Of those three, two galaxies 
(NGC 6339 \& UGC 11557, figure 4 \& 5) have signs of warps, and the remaining 
galaxy (UGC 11861, figure 6) has a bimodal H~I distribution (perhaps the sign 
of a central depression). 
The observed properties of the six galaxies are summarized in table 2.  
The main galaxies appear to be typical spiral galaxies, having rotation 
velocities between 100-180 km s$^{-1}$, dynamical masses between 
6.5-18$\times$10$^{10}$M$_{\odot}$, and H~I masses of 
1-7$\times$10$^9$M$_{\odot}$.  All six galaxies have R$_{HI}$/R$_{25}$ ratios 
(as measured at the 1 M$_{\odot}$/pc$^2$ level{\footnote{1 M$_{\odot}$/pc$^2$
= 1.2$\times$10$^{20}$cm$^{-2}$}) that are comparable with the average
value of 1.8$\pm$0.4 determined by Broeils (1992).  The galaxies appear to 
have a very extended, low column density component to the H~I; however, this 
may be due to the large angular size of our beams.  The M$_{HI}$/L$_B$ ratio 
of the six galaxies range from 0.46-4.2.  While the term ``gas-rich'' is not 
well-defined in the literature, it seems to be applied to galaxies with M/L as 
low as 0.24 depending on the morphological type of the galaxy (Huchtmeier 
1985) and appears to be uniformly applied to galaxies with M$_{HI}$/L$_B$ 
$\geq$ 0.4 (Young \& Lo 1997, Bothun \& Sullivan 1980).  Thus all our galaxies
qualify as gas-rich.  

Optically, these galaxies appear to have very little organized structure, with 
the notable exception of UGC 11124, which has a well-defined two-arm spiral 
pattern.  The stars in these galaxies are all found in regions of H~I high 
column densities ($\geq$5$\times$10$^{20}$ cm$^{-2}$), with no evidence to 
suggest that star formation has proceeded at lower column densities.  

The three galaxies with gas-rich companions or appendages are shown in 
figure 2; the lowest contour shown is at the 5$\sigma$ level 
(5$\times$10$^{18}$cm$^{-2}$).  Because the possible appendage on NGC 3246 is
not distinct from the primary galaxy, we can not tabulate its specific 
properties, and will not discuss it further in this paper.  The companions to 
UGC 9762 and UGC 11124 have properties (as listed in table 3) consistent with 
typical dwarf galaxies; they 
have obvious optical counterparts and rotational velocities of 10-50 km 
s$^{-1}$.  They are tens of kiloparsecs in size, and those two companions 
with optical magnitudes listed in NED{\footnote{The NASA/IPAC Extragalactic 
Database (NED) is operated by the Jet Propulsion Laboratory, California 
Institute of Technology, under contract with the National Aeronautics and 
Space Administration.}} have M$_{HI}$/L$_V$$\sim$ 1.0 indicating their 
gas-rich nature.  This assumes they are at 
the same distance as the galaxies they are near.  The optical counterparts of
UGC 11124N and UGC 9762S were previously cataloged as NPM1G+35.0421 and CG1284,
respectively.  The H~I to optical diameters of the companions are close to
those compiled Broeils (1992), and they also have an extended, low column 
density component.  The optical counterparts, whether compact or diffuse, show 
little distinctive structure, and are confined to the peak of the H~I emission.
We do not currently have optical redshifts for the companions, so we do not 
know whether the galaxies seen are actually associated with the H~I, but such 
a chance coincidence is highly unlikely.  

Using the maximum velocity along the major axis, and the
radius of that point in the formula: M$_{dyn}$=$\frac{V_{rot}^{2}R}{G}$, we 
can calculate the total enclosed mass within that radius for the companions.  
This is a lower limit since we do not know the inclination of these objects.  
The companions of UGC 11124 have small dynamical masses and seem to be 
dominated by 
the H~I.  Indeed H~I comprises 80\% of the mass of the southern companion to 
UGC 11124.  Unless these companions are nearly face-on, they must have less 
dark matter than is found in 
other dwarf galaxies.  For example, DDO 154 has M$_{HI}$/M$_{tot}$ of 7\% 
(Carignan \& Purton 1998), NGC 2915 of 4\% (Meurer {\it{et al.}} 1996), and 
DDO 168 \& DDO 105 of 10\% (Broeils 1992).  The companions
around UGC 9762 have more typical mass ratios as H~I makes up only a few 
percent of the total mass.  UGC 11124's companions also are less massive than 
those of UGC 9762.  In this small sample, we have been unable to find any 
properties that correlate with the presence or absence of gas-rich companions.

The projected separation between the primary galaxy and each companion is 
between 0.5 and 2 galaxy diameters in the H~I, equivalent to 25-90 kpc.  The 
projected velocity separations are 5-90 km s$^{-1}$.  Naturally we do not know
the true spatial or velocity separations for these systems.  This means that 
the companions may be interloping galaxies with large proper motions that are
not dynamically related to the primary galaxy.  Only with a large sample of 
galaxies can we determine statistically what fraction of apparent companions
are true companions.  

While we have chosen galaxies that are currently quiescent, we can not rule out
that these galaxies were not more active in the past.  The masses of the 
companions, however, allow us to say with reasonable certainty that these
objects could not be ejected by supernovae or O star winds.  Using the 
projected velocity separations and H~I masses of the companions, we 
calculate the kinetic energy of each companion; these values range between
10$^{53}$ to 10$^{55}$ ergs.  Compared to O stars, which put out 10$^{50}$ 
ergs over 10 Myr, or supernovae which put out 10$^{51}$ erg per 10 Myr
(Martin 1998), these clouds would require 10$^2$ - 10$^4$ supernovae 
in a small region and over a short time period to be ejected.  Furthermore, 
these clouds would then have to cool and form stars to become the galaxies
we see today.  While this is not impossible, it is unlikely that such an event 
occurred.  

Comparing our gas-rich companions with those found around other galaxies by
other authors, we find that they are somewhat similar objects.  Taylor
{\it{et al.}} (1996) detected companions around LSB dwarf galaxies; their 
companions have similar total and H~I masses to ours, but over twice the linear
size.  Furthermore, the LSB dwarf companions are more distant from the primary
galaxy, but these companions have roughly the same masses as their primary 
galaxies.  The H~I mass ratios of the H~II galaxies and companions studied by 
Taylor {\it{et al.}} (1995) are close to those of the UGC 9762 system;  the 
H~I and total masses are also similar.  The sizes of the H~II galaxy 
companions are bigger than our detections, but the companions have similar 
relative and linear separations.  The H~II galaxies in general, however, are 
smaller than those in 
our sample.  Compared to those companions found by Wilcots {\it{et al.}} 
(1996), our companions have similar H~I sizes, but are $\sim$10$\times$ more 
massive.  Overall, our companions are similar in mass to those found by Taylor 
{\it{et al.}} (1995,1996), but with smaller sizes.  The main difference between
our sample and that of Taylor {\it{et al.}} is the nature of the primary 
galaxy; ours are generally bigger late-type spirals and, hence, more similar 
to those of Wilcots {\it{et al.}} (1996), while Taylor {\it{et al.}} studied
dwarf galaxies.

\section{Discussion}

Using projected values, we can make an estimate of the effect these companions
have on their host galaxies and the likely fate of these triple systems.
The relevant values are summarized in table 4.  We have characterized the 
current strength of the interactions in these triple systems using the Dahari 
parameter:  
Q$_D$ = (M$_{companion}$/M$_{galaxy}$)/(R$_{sep}$/R$_{galaxy}$)$^3$ 
(Dahari 1984, Byrd {\it{et al.}} 1986).  The Dahari parameter uses the mass
ratio and relative separation within an interacting system to quantify
the strength of the interactions occurring.  For the UGC 11124 \& UGC 9762 
systems the values of Q$_D$ are between 6$\times$10$^{-5}$ and 
6$\times$10$^{-3}$.  As
compared to ``strong'' (Q$_D\geq$1) interactions which could trigger 
Seyfert activity (Dahari 1984), or even minor mergers with Q$_D\sim$0.1 (e.g. 
Hernquist \& Mihos 1995), the 
interactions within these triple systems are quite weak.  For example, the 
Q$_D$'s are comparable to those of Taylor {\it{et al.}}'s (1995,1996) and 
Wilcots {\it{et al.}}'s (1996) samples.  The companions, therefore,
currently have very little effect of the primary galaxies.  This absence of a 
triggering mechanism could help explain the lack of star formation in the main 
galaxies that we would otherwise expect during a tidal interaction.  

Even though the current interactions are quite weak, the eventual decay and 
collapse of these triple systems should cause more noticeable effects; but when
will such a collapse occur?  We can calculate the dynamical friction timescale
using formulas 7-26 \& 7-13b from Binney \& Tremaine (1987): 
\begin{equation}
t_{fric}=\frac{2.5{\times}10^{14}{\times} R^2 V}{M_{comp}{\times}ln(\Lambda)} 
yr
\end{equation}
where
\begin{equation}
\Lambda = \frac{RV^2}{M_{comp}G}
\end{equation}
for R is the separation between centers in kpc, V is the velocity difference 
in km s$^{-1}$ between the companion and the primary galaxy, and M$_{comp}$ 
is the mass of the companion in solar 
masses.  The dynamical friction timescale characterizes the amount of time it
will take a companion to spiral into the center of the primary galaxy provided
that companion is orbiting within the dark matter halo of the primary galaxy.
Simulations of triple systems of galaxies suggest that they will survive a few
orbit times, making this another good measure of the lifetime of such a group 
(Barnes 1989).

Using the projected values of R and V, we find dynamical friction times of 
about 6 Gyr for UGC 11124N \& UGC 9762S, and 136 Gyr for UGC 11124S.  We can
not calculate a physical timescale for UGC 9762N.  While these times are long, 
they are comparable to the orbit times of the companions around the primary 
galaxy (simply 2$\pi$R/V).  Zaritsky 
{\it{et al.}} (1997) found that satellites are preferentially in 
orbits that lie in the plane of the primary galaxy and are in circular 
rotation.  Assuming that our companions follow this trend, we can calculate
their rotation velocity around the primary galaxy and their true separation.  
Because it is the difference in velocity
between the companion and the dark matter halo of the primary galaxy which 
provides dynamical friction, we assume that the halo is rotating with the same
velocity as the last measured point of the H~I disk.  Recalculating t$_{fric}$
and t$_{orbit}$ for this scenario, we find timescales of $\sim$5-15 Gyr.  
These numbers suggest that these triple galaxy systems are quite stable.  

Comparing t$_{fric}$ for our systems with those for other satellites found by
other authors, we find they are rather similar.  Taylor {\it{et al.}} 
(1995,1996) found companions around LSB and H~II dwarf galaxies.  Using the
average projected separations from these studies these companions have 
dynamical friction timescales of $\sim$5 \& 30 Gyr, respectively, and orbit 
times of 8 \& 17 Gyr.  For the
SBm's with companions observed by Wilcots {\it{et al.}} (1996), t$_{fric}$
\& t$_{orbit}$ $\sim$1-6 Gyr.  Finally, for the median satellite galaxy 
in the sample of Zaritsky {\it{et al.}} (1997) we derive a t$_{orbit}$ of 
$\sim$ 30 Gyr.  From these values we can see that our companions have an 
expected longevity quite similar to the satellite galaxies studied by
Taylor {\it{et al.}} (1995,1996), shorter than Zaritsky {\it{et al.}} (1997), 
but longer than those found by Wilcots {\it{et al.}} (1996).  

Because the dynamical friction formalism is derived for stellar systems, it 
is possible that other processes, such as gas drag, may cause these
systems to collapse sooner or be torn apart.   There is an indication that this
is happening in the case of the northern companion to UGC 11124, which is 
currently being accreted.  

In the case of UGC 11124N, the companion has 3\% of the H~I mass and
0.5\% of the total mass of UGC 11124.  As this companion is accreted by 
UGC 11124 it is unlikely that it will dramatically alter the structure of
the primary galaxy.  On the other hand, such a merger will result in a 
substantial infusion of gas into UGC 11124, which may be enough to 
significantly alter the future star formation of the galaxy.  Hernquist \&
Mihos (1995) have investigated the effects of ``minor mergers'' where the mass
ratio involved is 1:10 and find that these events can trigger bar formation and
central activity as well as general star formation in galaxies.  As the 
interaction of UGC 11124N with UGC 11124 involves smaller mass ratios than 
studied by Hernquist \& Mihos the effects are likely to be less dramatic.

Are these gas-rich companions a sign of ongoing galaxy assembly?  These 
systems have some interesting similarities to those
seen in cold dark matter galaxy formation simulations which suggests that we 
may be seeing the continuing assembly of these galaxies' disks.  Simulations by
Navarro, Frenk, \& White (1995) of isolated galaxy formation find small gaseous
satellites that have yet to be accreted by z=0.  This is also found in earlier 
simulations by Katz (1992).  These simulations find ``one or two small 
satellites'' that contain about 10\% of the primary galaxy's disk mass which 
represent the gas recently accreted by the galaxy's halo (Navarro, Frenk, \& 
White 1994).  The similarity of these simulations to the cases of UGC 9762 \&
UGC 11124 suggest that we may be seeing the end of the mass assembly 
process.  The accretion of UGC 11124N by UGC 11124 may well be an example of 
ongoing mass assembly.  Overall, the process
of galaxy assembly will take over a Hubble time to complete if these systems
are indicative of the process.

While UGC 11124 suggests that the mass assembly process could be 
continuing into the present day, the remaining galaxies in our
sample do not show such obvious signs.  This could mean that galaxy assembly
has finished for most galaxies, or that assembly is continuing with smaller 
masses than we can detect.  Because of the small size of our sample we 
can not yet make a statement as to the current state of galaxy assembly for
most galaxies.

\section {Conclusions}

We have observed six extremely isolated galaxies in a search for H~I remnants
of the initial gaseous reservoir out of which these galaxies formed.  Two of 
the ``isolated'' galaxies, UGC 9762 and UGC 11124, turn out to be the 
primary component of a triple galaxy system.  In each case the secondary and
tertiary galaxies are faint, gas-rich dwarfs with dynamical masses of about 
0.5-10\% of the primary member.  These triple systems have small velocity 
dispersions and are no more than 90 kpc in radius.  While these systems appear
quite stable with dynamical friction timescales of greater than 5 Gyr, there
is at least one case of merging in these systems.  UGC 11124 is 
slowly accreting its dwarf galaxy companion, UGC 11124N.  This event will 
increase the gas mass of UGC 11124 by 3\% possibly triggering an episode of 
star formation.  Both the UGC 9762 \& UGC 11124 systems bear a resemblance to 
those seen in CDM galaxy formation simulations, suggesting that we are seeing 
the end of the mass assembly process and an important galaxy evolution driver 
in these two systems.

\acknowledgements

Special thanks to Dr. Liese van Zee who provided valuable assistance with the 
H~I reductions in Socorro, and was kind enough to observe UGC 9762 and UGC 
11124 during an observing run on Kitt Peak.  Thanks also to all the staff at 
the VLA
for their help and hospitality during data reduction.  Thanks to Linda Sparke
for reading this manuscript and correcting errors in our dynamical friction
calculations.  The authors would also
like to thank the anonymous referee for prompt and helpful comments which
improved this paper.  This research has made use of the NASA/IPAC
Extragalactic Database (NED) which is operated by the Jet Propulsion 
Laboratory, California Institute of Technology, under contract with the 
National Aeronautics and Space Administration.  The Digitized Sky Surveys were
produced at the Space Telescope Science Institute under U.S. Government
grant NAG W-2166. The images of these surveys are based on photographic data 
obtained using the Oschin Schmidt Telescope on Palomar Mountain and the UK 
Schmidt Telescope. The plates were processed into the present compressed 
digital form with the permission of these institutions. 
The National Geographic Society - Palomar Observatory Sky Atlas (POSS-I) was 
made by the California
Institute of Technology with grants from the National Geographic Society.
This work was partially supported by NSF grant AST 96-16907 to EMW.

\vfil

\centerline {\bf{Figure Captions}}

\figcaption{}UGC 11124 and companions.  The left panel has total H~I contours
on a B band 0.9m KPNO images.  Contours are at 0.5, 1, 5, 10, 50, 
100$\times$10$^{19}$cm$^{-2}$.  The right panel has the velocity field 
overlaid on the total intensity map.  Velocity contours are every 15 km 
s$^{-1}$.

\figcaption{}Same as figure 1, but for UGC 9762.  Velocity contours are every
20 km s$^{-1}$.

\figcaption{}NGC 3246.  The left panel has total H~I intensity on a DSS image.
Contours are at 0.5, 1, 5, 10, 50, 100$\times$10$^{19}$cm$^{-2}$.  
The right panel has the velocity field overlaid on the total intensity map.  
Velocity contours are every 20 km s$^{-1}$.   

\figcaption{}same as figure 3, but for NGC 6339.

\figcaption{}same as figure 3, but for UGC 11557.

\figcaption{}same as figure 3, but for UGC 11861.

\end{document}